\documentclass{elsart}

\usepackage{graphicx}

\usepackage{lineno}
\usepackage{appendix}

\usepackage{amssymb}
\usepackage{amsmath}
\usepackage{pxfonts}
\usepackage{footnote}
\usepackage{multirow}
\usepackage{varwidth}
\usepackage{array}
\usepackage{epstopdf}
\usepackage{url}

\journal{}

\begin{document}

\thispagestyle{empty}
\begin{Large}
	\textbf{DEUTSCHES ELEKTRONEN-SYNCHROTRON}
	
	\textbf{\large{Ein Forschungszentrum der Helmholtz-Gemeinschaft}\\}
\end{Large}

DESY-25-013

January 2025

\begin{eqnarray}
\nonumber &&\cr \nonumber && \cr \nonumber &&\cr
\end{eqnarray}
\begin{eqnarray}
\nonumber
\end{eqnarray}
\begin{center}
	\begin{Large}
		\textbf{A Femtosecond X-ray Undulator Source for EUV Pump/X-ray Probe Experiments at FLASH2}
	\end{Large}
	\begin{eqnarray}
	\nonumber &&\cr \nonumber && \cr
	\end{eqnarray}
	
	%
	
	R. Khubbutdinov, V. Kocharyan, E. Saldin
	%
	\textsl{\\Deutsches Elektronen-Synchrotron DESY, Hamburg}
	\begin{eqnarray}
	\nonumber
	\end{eqnarray}
	\begin{eqnarray}
	\nonumber
	\end{eqnarray}
	%
	\begin{eqnarray}
	\nonumber
	\end{eqnarray}
	\begin{large}
		\textbf{NOTKESTRASSE 85 - 22607 HAMBURG}
	\end{large}
\end{center}
\clearpage
\newpage

\begin{frontmatter}



\title{A Femtosecond X-ray Undulator Source for EUV Pump/X-ray Probe Experiments at FLASH2}


\author[DESY]{R. Khubbutdinov,}
\author[DESY]{V. Kocharyan,}
\author[DESY]{E. Saldin}
\address[DESY]{Deutsches Elektronen-Synchrotron DESY, Hamburg, Germany}

\begin{abstract}

The recently implemented APPLE-III device at FLASH2 is an undulator source providing tunable, femtosecond spontaneous X-rays in the range of 2 - 3.5 keV at the 3rd and 5th harmonics. This work demonstrates its applicability for EUV FEL laser pump/probe absorption and diffraction experiments.

\end{abstract}

%
%

%
\end{frontmatter}



\section{ Introduction }

In the course of the FLASH2020 project at DESY \cite{F2}, an APPLE-III undulator was developed and installed as a (third harmonic) afterburner at the current FLASH2 FEL \cite{T}. It is used to generate circularly polarized radiation with a wavelength of 1.3 nm to 1.77 nm for the investigation of the L-edges of Fe, Co and Ni.  The use of circularly polarized soft X-ray radiation at the FLASH-FEL is a very versatile tool for the investigation of dynamic properties in nanomagnetism studies. 

The range of applications of the APPLE III undulator at FLASH2 is wider than one might think. It is natural to take advantage of this insertion device to provide spontaneous X-ray undulator radiation. The elliptically polarized APPLE-III undulator allows not only the production of circularly polarized radiation, but also an arbitrary polarization mode. Of particular interest is the use of the high harmonic generation option in the linear (i.e. planar undulator) mode.

One of the most important methods used at the XFEL sources for time-resolved studies of atoms, molecules, and solid-state systems is the pump-probe technique, in which an optical laser pulse initiates dynamics that are later probed by an X-ray pulse. To capture the dynamics, the X-ray probe measurement can employ various methods including scattering, diffraction, emission, X-ray near-edge absorption, etc. \cite{PP}.    
Here, we propose an approach using spontaneous emission at the 3rd and 5th harmonic of the APPLE-III undulator to drive the EUV pump/X-ray probe experiments at FLASH2. 

\section{The Existing Scheme of Time-Resolved X-ray Absorption Spectroscopy}

One of the schemes being developed to extract femtosecond X-ray pulses from a storage ring is called a slicing source. 
The slicing method is based on the resonant energy exchange between a femtosecond laser pulse, whose electric field can achieve very high value and temporally short "slices" of electrons within a few tens of picoseconds long electron bunch. The femtosecond laser pulse and the original electron bunch co-propagate in the insertion device of a storage ring, while the electric field of the laser pulse modulates the energy spread of a temporally small portion of the electron bunch.

The electrons interacting with the laser pulse will gain a significant energy dispersion.   
The modulation of the electron energy leads to a spatial separation from the main bunch by means of a bending magnet in the storage ring and is then used to generate femtosecond X-rays at another bending magnet or wiggler. The background contribution from the remaining electrons in the long bunch can be reduced by placing an aperture at the image plane of the source to select only the short X-ray pulses originating from the sliced electrons. This radiation has approximately the same duration as the femtosecond laser pulse. The slicing concept provides femtosecond X-ray pulses that are naturally synchronized with the optical laser. Currently, the repetition frequency is on the order of a few kHz. The overall flux is on the order of ~$10^6$ photons/second. 

This technique was first demonstrated at the ALS at LBL in the US. Besides the ALS, the slicing technique has been successfully implemented in Germany (BESSY II) and Switzerland (SLS). The laser slicing facility at BESSY II delivers about 100 fs soft X-ray pulses in the 250 - 1400 eV range with an average flux of about $10^6$ photons/s in a $10^{-3}$ bandwidth. Femtosecond pulses from a hard (5-12 keV) X-ray source have been demonstrated at the SLS, providing the order of $10^6$ photons/s in a $10^{-2}$ bandwidth \cite{SH}.

\section{Operation of an APPLE-III Undulator at 3rd and 5th Harmonics }

This section describes the spectral and angular density of the undulator radiation source at the arbitrary harmonic and the total photon flux produced by a filament electron beam within the central cone. In particular, we are interested in the operation of the APPLE-III undulator source and the output parameters of the X-ray beam at the 3rd and 5th harmonics of the device. Calculations are presented for the case of the APPLE-III undulator at FLASH. The problem of the undulator magnetic field errors and the influence of the electron beam parameters on the radiation properties of the photon beam are discussed below.  

Let us consider the basic parameters of planar undulator radiation. The undulator equation
\begin{equation} \label{undi0}
	\omega = \frac{2ck_w\gamma^2}{1 +K^2/2 +\gamma^2\theta^2},
\end{equation} 
tell us the frequency of radiation at the fundamental harmonic as a function of undulator period $\lambda_w = 2\pi k_w$, undulator parameter $K$, electron energy $mc^2\gamma$, and polar angle $\theta$. Note that for radiation within the cone of half angle 
\begin{equation} \label{undi1}
	\theta_c = \frac{\sqrt{1 + K^2/2}}{\gamma\sqrt{N_w}},
\end{equation} 
the relative spectral bandwidth is 
\begin{equation} \label{undi1a}
	\Delta\omega/\omega = 0.88/N_w,
\end{equation} 
where $N_w$ is the number of undulator periods. The spectral and angular density of the radiation energy emitted by a single electron during the undulator pass is given by the expression 
\begin{equation} \label{undi2}
	\frac{d^2 E}{d\omega d\Omega} = \frac{e^2 N_w^2\gamma^2A_{JJ}^2K^2}{2c(1 +K^2/2)^2}\frac{\sin^2[\pi N_w(\omega - \omega_0)/\omega_0]}{[\pi N_w(\omega - \omega_0)/\omega_0]}.
\end{equation} 
Here $\omega_0 = 2\gamma^2k_w/(1 +K^2/2)$ is the resonance frequency, $A_{JJ} =J_0(Q) - J_1(Q)$, where $J_k$  is the Bessel function of $k$th order, and $Q = K^2/(4 + 2K^2)$. 

Now we would like to calculate the energy radiated into the central cone. In the small angle approximation the solid angle is equal to $d\Omega = \theta d\theta d\phi$. Integration of spectral and angular density over $\omega$ and $\phi$ gives us factors $\omega_0/N_w$ and $2\pi$, respectively. We also have to integrate over $\theta$ from $0$ to $\theta_c$. Thus, the energy radiated into the central cone by a single electron is given by 
\begin{equation} \label{undi3}
	\Delta E_c = \frac{\pi e^2A_{JJ}^2\omega_0K^2}{c(1 +K^2/2)}.
\end{equation} 
To complete the calculation of the undulator characteristics at the fundamental harmonic an expression is needed for the photon flux within the central cone.  We divide the radiation energy by the energy per photon and obtain
\begin{equation} \label{undi4}
	\frac{dN_p}{dt} = \frac{\pi\alpha A_{JJ}^2K^2N_ef}{1 +K^2/2},
\end{equation}  
where $\alpha = 1/137$ is fine structure constant, $N_e$ is a number of electrons in a bunch, $f$ is the bunch repetition rate. This result is obtained in the framework of a filament electron beam model.

We now turn to the operation of an undulator at high harmonics. One of the specific properties of a planar undulator is the ability to produce radiation fields at odd harmonics that are highly directional along the undulator axis.
The emission on the axis of the even harmonics vanishes \cite{O}. At resonance, a single electron passing a planar undulator radiates an electromagnetic wave with $nN_w$ cycles, where $n = 1, 3, 5, ...$ is the harmonic number. For the radiation within central cone of half angle
\begin{equation} \label{undi4a}
	\theta_c = \frac{\sqrt{1 + K^2/2}}{\gamma\sqrt{nN_w}},
\end{equation} 
the relative spectral FWHM bandwidth is $\Delta\omega/\omega = 0.88/(nN_w)$ . The spectral and angular density of the radiation energy emitted by a single electron during the undulator pass is given by the expression 
\begin{equation} \label{undi5}
	\frac{d^2 E}{(d\omega d\Omega)} = \frac{e^2 n^2N_w^2\gamma^2(A^n_{JJ})^2K^2}{2c(1 +K^2/2)^2}\frac{\sin^2[\pi nN_w(\omega - \omega_0)/\omega_0]}{[\pi nN_w(\omega - \omega_0)/\omega_0]},
\end{equation} 
where $\omega_0 = 2n\gamma^2k_w/(1 +K^2/2)$ is the resonance frequency, 
$A^n_{JJ} =J_{(n-1)/2}(Q) - J_{(n+1)/2}(Q)$, and $Q = nK^2/(4 + 2K^2)$. The energy radiated into a central cone by a single electron in the case of an arbitrary harmonic number is given by
\begin{equation} \label{undi6}
	\Delta E_c = \frac{\pi e^2(A^n_{JJ})^2n\omega_0K^2}{[c(1 +K^2/2)]}
\end{equation} 
The photon flux produced by a filament electron beam within the central cone in a planar undulator for the $n$th harmonic is given by. 
\begin{equation} \label{undi7}
	\frac{dN_p}{dt} = \frac{\pi\alpha (A^n_{JJ})^2K^2N_ef}{(1 +K^2/2)}.
\end{equation} 
We performed calculations for the harmonics $n =1, 3$ and 5.  
For the spontaneous photon flux at the exit of APPLE-III undulator within the central cone on the 1st, 3rd and 5th harmonic, we find   $dN_p/dt \simeq 6\cdot 10^{10}$/sec,  2.4$\cdot 10^9$/sec, and $2 \cdot 10^8$/sec respectively.
The undulator and beam parameters used in the calculations are as follows: the undulator period length is 17.5 mm, the $K$ value is about 0.91, the total undulator length is 2.5 meters, the electron beam energy (max) is 1.35 GeV, the electron energy spread is about 0.5 MeV, the normalized emittance is 1.4 mm$\cdot$mrad in both transverse directions, the beta function is 6 meters, the bunch charge (max) is assumed to be 1 nC, the number of bunches in the macro-pulse is 100, and the repetition rate is about 10 Hz.  

There are several points to be made about the above results. One interesting question is about the magnetic field errors of the APPLE III undulator.  As can be seen from the radiation emission formula, the effect of the field errors is pronounced in the phase term, mainly introducing the "phase noise" which can reduce the angular and spectral flux density \cite{W}. The magnetic measurements of the APPLE-III undulator revealed that rms phase error is about 3 degrees. 
The influence of the phase errors on the radiation properties of the 5th harmonic should still be investigated. The conclusion of studies presented in \cite{W} is that an rms error of about 3 degrees is good enough for high harmonic operation.  For example, the reduction in intensity on the 5th harmonic is only a few percent.

The results presented above refer to the case of filament electron beam.
This is a very useful approach, allowing one to study many features of the undulator radiation by means of relatively simple tools. However, it is relevant to make some remarks on the applicability of this beam model at FLASH. 
If there is an electron energy(rms) spread within the bunch, $\Delta\gamma/\gamma$, there will be a corresponding photon energy spread given by $\Delta E_p/E_p = 2\Delta\gamma/\gamma$. If there is an (rms) angular divergence $\sigma'$ within the bunch, there will be a corresponding photon energy spread given by
\begin{equation} \label{undi8}
	\frac{\Delta E_p}{E_p} = \frac{\gamma^2(\sigma')^2}{(1 + K^2/2)}.
\end{equation} 
In what follows we use the following assumption: $4\Delta\gamma/\gamma \ll 1/(nN_w)$, $(\sigma')^2  \ll (1 + K^2/2)/(\gamma^2nN_w)$. When these conditions are satisfied, the energy spread and the angular divergence cause a spectral broadening less than $1/(nN_w)$ and the central cone will be rather well defined in terms of both its angular distribution and spectrum.     

Estimating the energy spread and emittance effects are the next questions to be investigated. If we consider the data 
for the undulator and beam parameters presented above, we find that we can neglect the influence of the beam distributions in the case of radiation at the fundamental harmonic and treat the setup as an undulator source with a filament electron beam. On the other hand, the theoretical analysis performed above indicates that there is an influence of the angular and energy distributions in the case of higher harmonic radiation. 
A quantitative answer to the parameters of the photon distributions for the 3rd and 5th harmonics can be found from numerical simulations. We will discuss this topic further in the section \ref{Numerical}.

\section{X-ray Beam Transport}

The optical system that transports EUV photons from the FLASH2 FEL undulator to the experimental hall is located downstream of the APPLE-III afterburner. The beam transport path from the source point of the FEL radiation, located within the last segment of the FEL undulator, to the scientific instruments in the experimental hall is 90 meters long.  
The only optics encountered in the FLASH2 photon beamline are a pair of plane-reflecting offset mirrors that shift the beam by a 50 mm from the optical axis. These mirrors remove the high-energy Bremsstrahlung radiation that co-propagates with the FEL radiation and are an essential part of the radiation safety of all instruments. The incident mirror angle is fixed at one degree. The offset mirrors are installed 20 meters downstream of the FEL undulator source. 

It is important to calculate the expected overall efficiency of the transport photon beamline. The combined effect of two fixed-angle offset mirrors cannot be neglected. The reflectivity of the mirrors has been calculated using the CXRO code \cite{CXRO}. The mirrors are assumed to be platinum coated. In the photon energy range of 2-3 keV, platinum has a reflectivity of about 0.5 at a grazing angle of one degree. The overall efficiency of the transport beamline is on the order of 0.25.     

The separation of the pump and probe pulses can be achieved by using the existing magnetic corrector (weak dipole magnet) installed between the FEL and APPLE-III undulators.
Thus, a sufficient separation can be achieved by kicking an electron beam at a relatively small angle of about 0.1 mrad. It should be noted that we consider a spontaneous emission. As such, the impact of this kick on the performance of the X-ray source is negligible. The transverse shift of the X-ray beam caused by the use of a kick is about one centimeter in the experimental hall.

\section{Numerical Simulations of Spontaneous Emission}\label{Numerical}

In the previous sections, a preliminary study of the X-ray source was performed using an ideal (i.e. filament and monoenergetic) electron beam approximation, which in a sense resembles a perfect undulator source. In the next stage of our analysis, a quantitative description of the X-ray source was obtained using numerical simulation. In order to have a more realistic case, the X-ray source was modeled taking into account all electron beam distributions. The radiation parameters of APPLE-III were simulated with the code XRT \cite{xrt}. 

\begin{figure}[tb]
	\centering
	\includegraphics[width=1.0\textwidth]{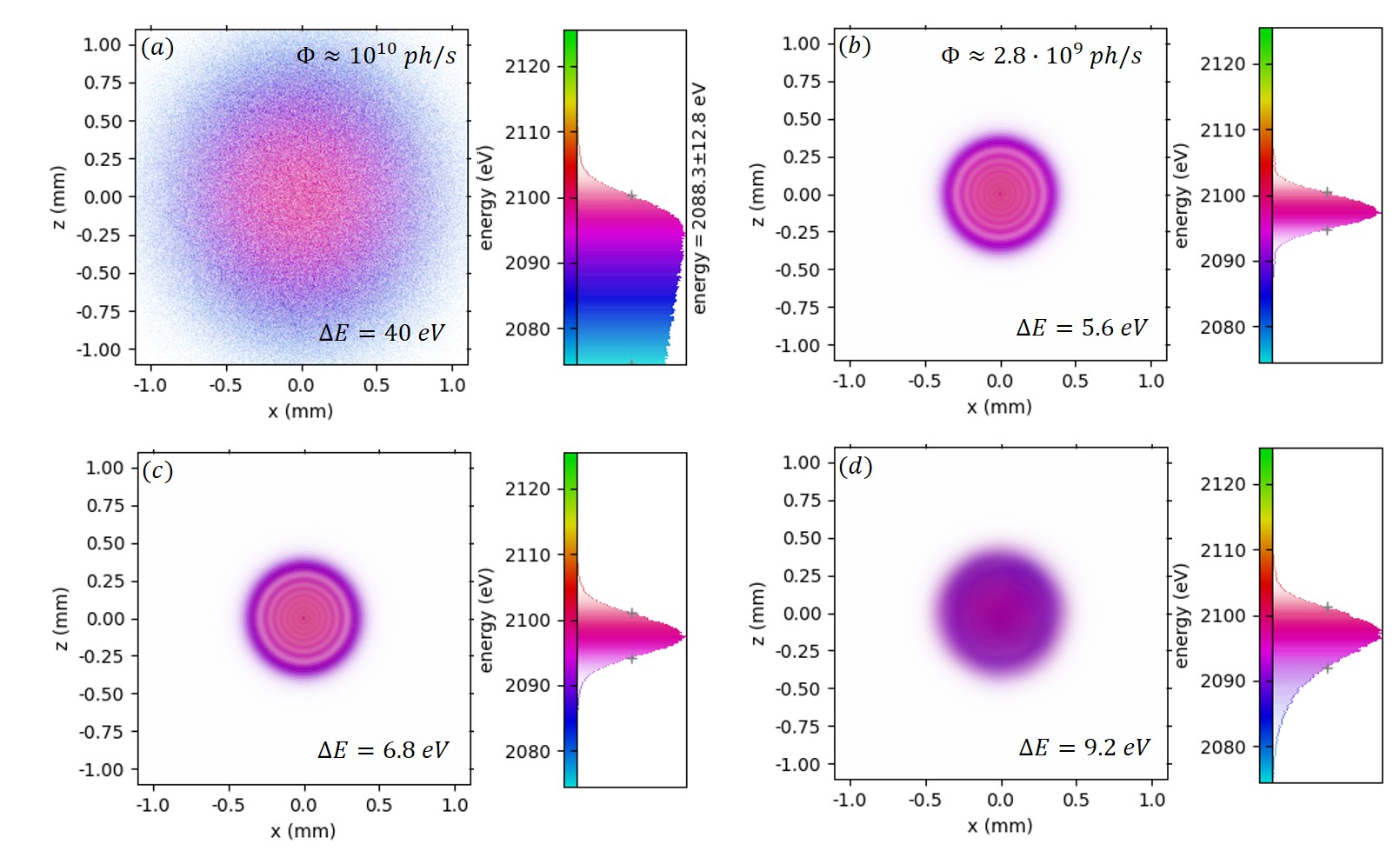}
	\caption{Intensity distribution of the source near the 3rd harmonic in the 40 eV energy range (a), in the central cone for the filament beam without energy spread (b), for the filament beam with energy spread of 4$\cdot 10^{-4}$ (c), for the electron beam with normalized emittance of 1.4 mm$\cdot$mrad and the energy spread of 4$\cdot 10^{-4}$.} \label{Fig1}
\end{figure}

\begin{figure}[tb]
	\centering
	\includegraphics[width=1.0\textwidth]{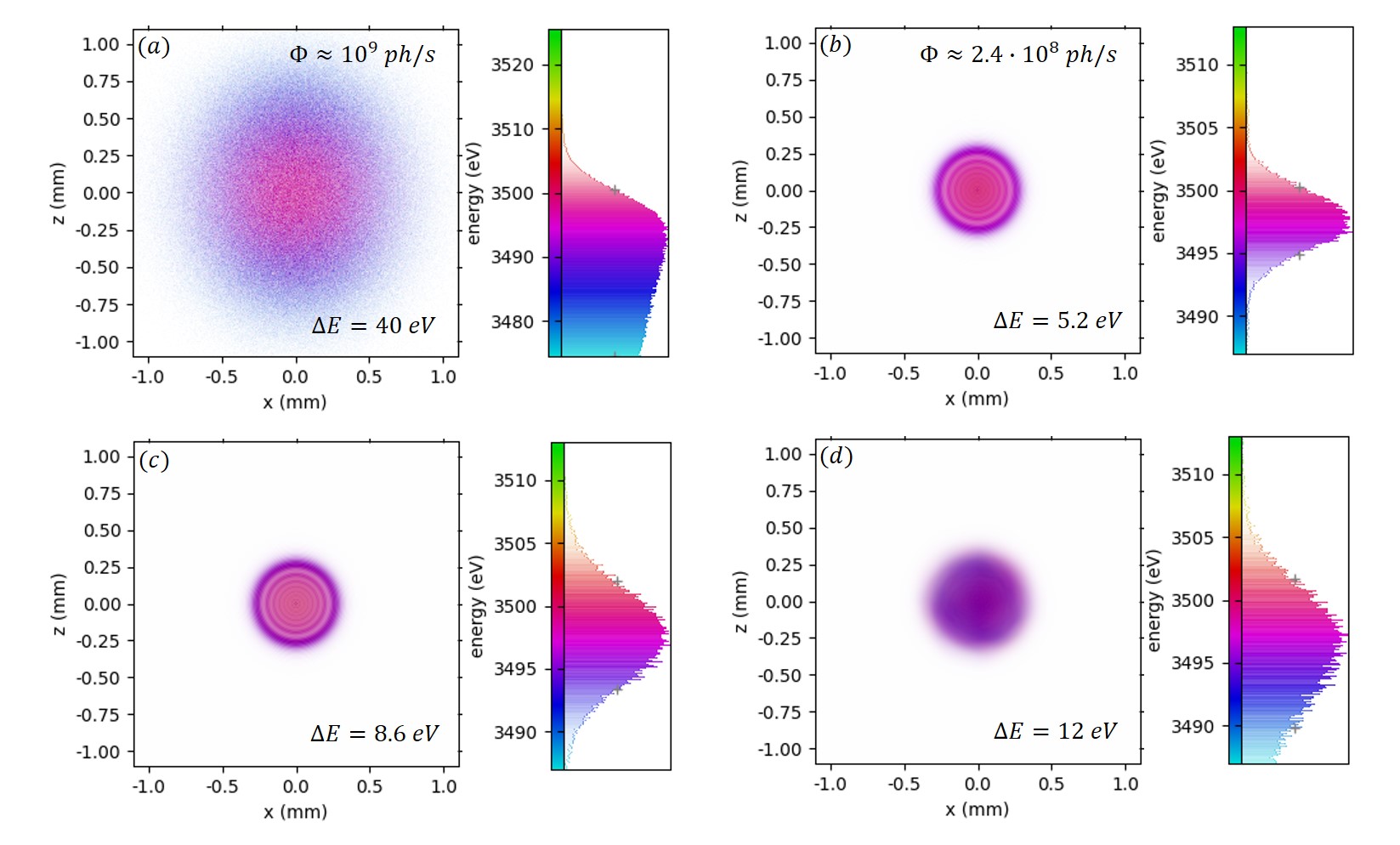}
	\caption{Intensity distribution of the source near the 5th harmonic in the 40 eV energy range (a), in the central cone for the filament beam without energy spread (b), for the filament beam with energy spread of 4$\cdot 10^{-4}$ (c), for the electron beam with normalized emittance of 1.4 mm$\cdot$mrad and the energy spread of 4$\cdot 10^{-4}$.} \label{Fig2}
\end{figure}

The X-ray source optical system consists of an APPLE-III undulator, offset mirrors, and an angular pinhole filter. The angular pinhole filter, located after the APPLE-III undulator, is used to narrow the angular distribution of the transmitted X-rays (down to the central cone). The results of simulations performed with the XRT code are shown in Fig. \ref{Fig1} - \ref{Fig2}. As can be seen from the figures, the photon flux within the central cone is almost the same as in the case of the analytical model. However, a complete, straightforward analysis of the filtered photon spectrum can only be performed numerically.  

\section{FLASH2 FEL Simulations: Spent Electron Beam Properties}

Let us discuss the important issue of the influence of the FLASH2 FEL on the performance of the X-ray source.
First of all, it has to be said that the FEL process does not affect the angular distribution of the electron beam.
According to XFEL physics, the FEL process is only capable of altering the longitudinal motion of the electrons, but not the betatron motion. As a consequence, the FEL process does not influence the angular distribution of the spontaneous emission of the spent electron beam in the APPLE-III undulator. From the above analysis, it follows that the number of photons radiated within the central cone does not show any sensitivity to the FEL operation.  Consequently, the question arises whether the FEL effects play a role  and whether the FEL process has to be taken into account in our case of interest.  In this regard, knowledge of the FEL-induced electron energy distribution along the electron bunch at the exit of the FLASH2 FEL is still required to calculate the X-ray photon energy spread within the central cone.

We consider a simple case when the FEL is tuned in a deep linear regime. In this case, we have to deal with the initial energy distribution along the electron bunch. In the following, we will consider only one setting of the FLASH2 FEL at 250 pC electron bunch charge. The setup for a maximum charge of 1 nC is still possible, but requires further adjustments concerning electron beam optics. 

\begin{figure}[tb]
	\centering
	\includegraphics[width=0.7\textwidth]{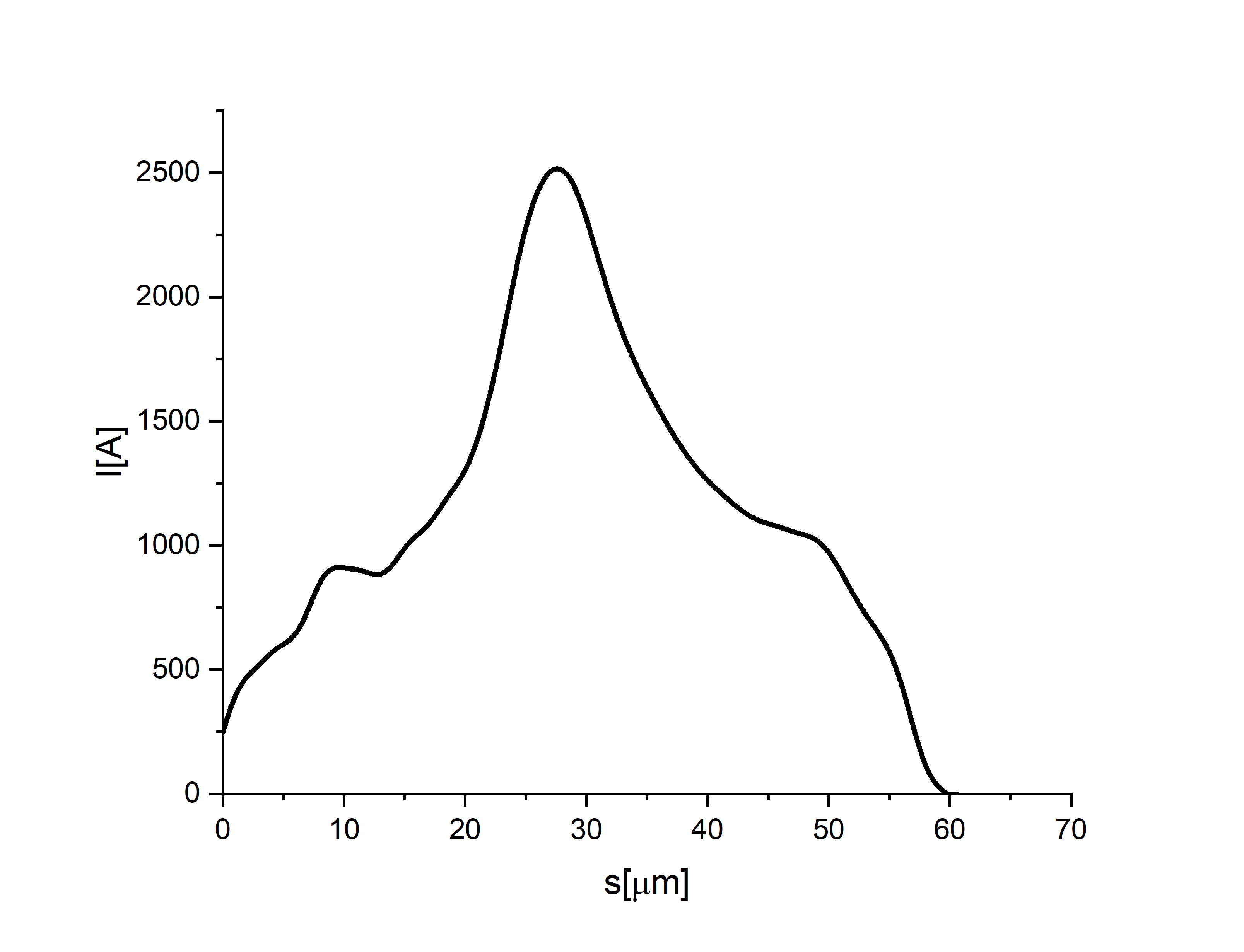}
	\caption{Result from electron beam start to end simulations at the entrance of EUV FEL undulator. Current profile.  } \label{Current}
\end{figure}

\begin{figure}[tb]
	\centering
	\includegraphics[width=0.7\textwidth]{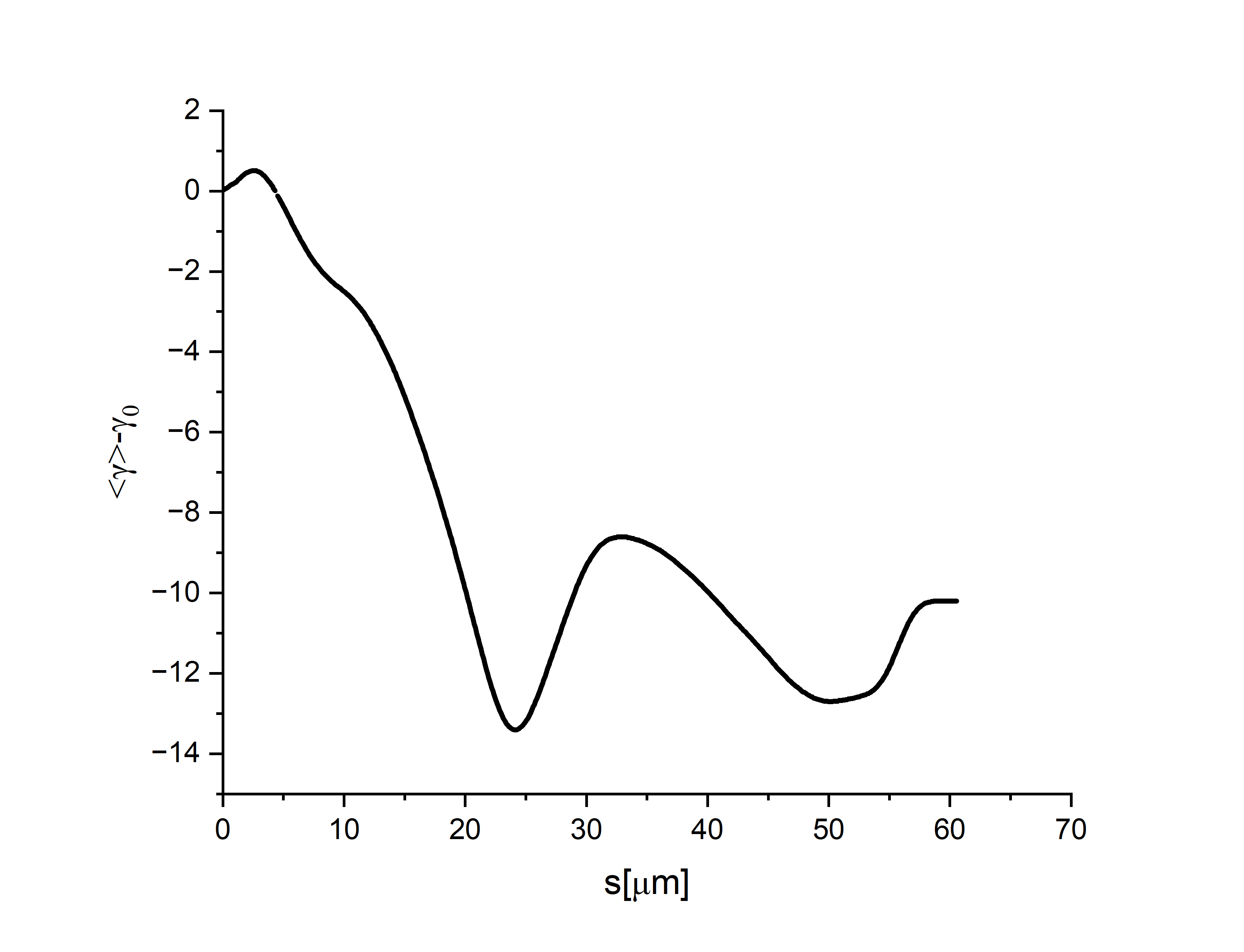}
	\caption{Result from electron beam start to end simulations at the entrance of EUV FEL undulator. Energy profile along the bunch.} \label{Eng}
\end{figure}

\begin{figure}[tb]
	\centering
	\includegraphics[width=0.7\textwidth]{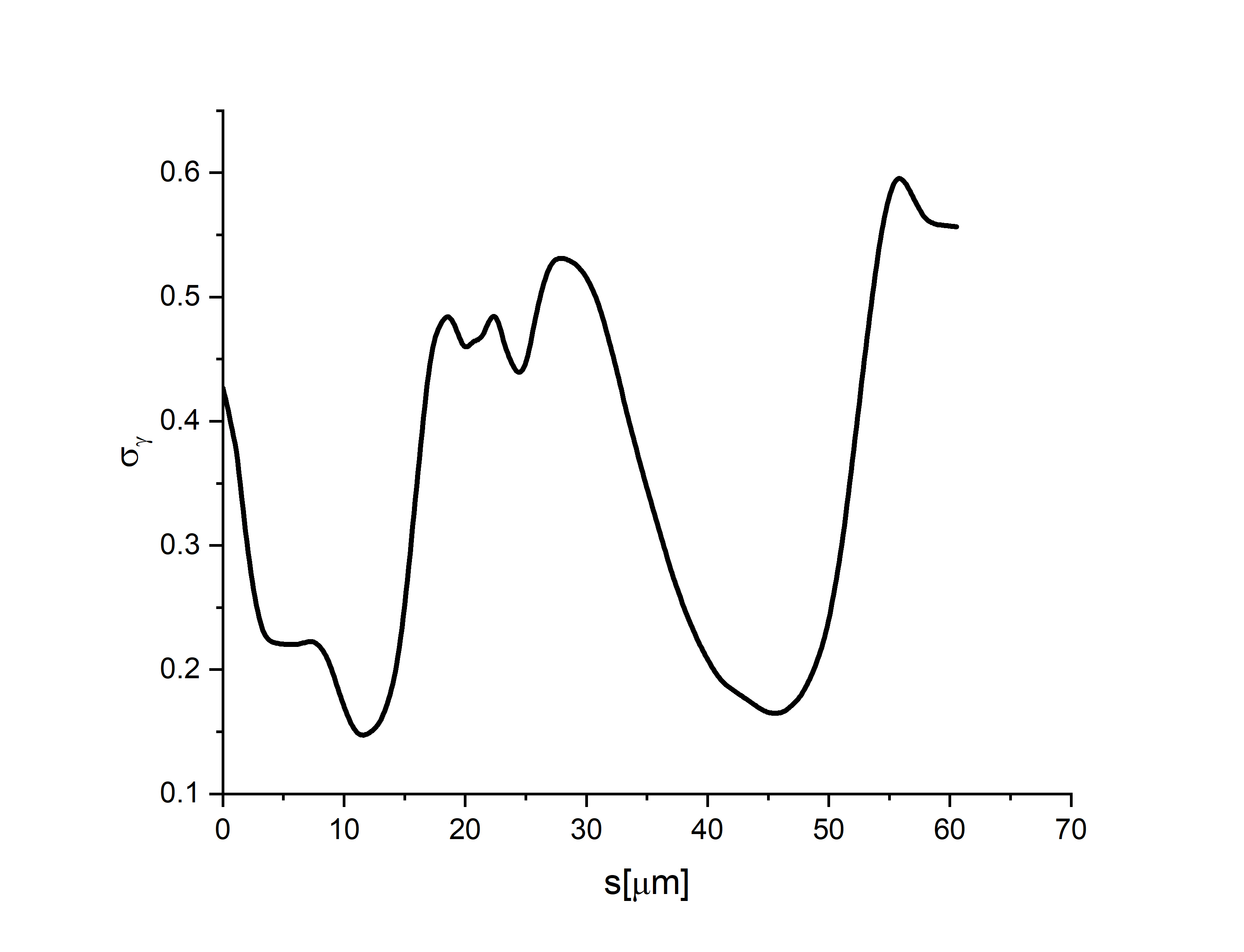}
	\caption{Result from electron beam start to end simulations at the entrance of EUV FEL undulator. Electron beam energy spread profile.} \label{EngSp}
\end{figure}

\begin{figure}[tb]
	\centering
	\includegraphics[width=0.7\textwidth]{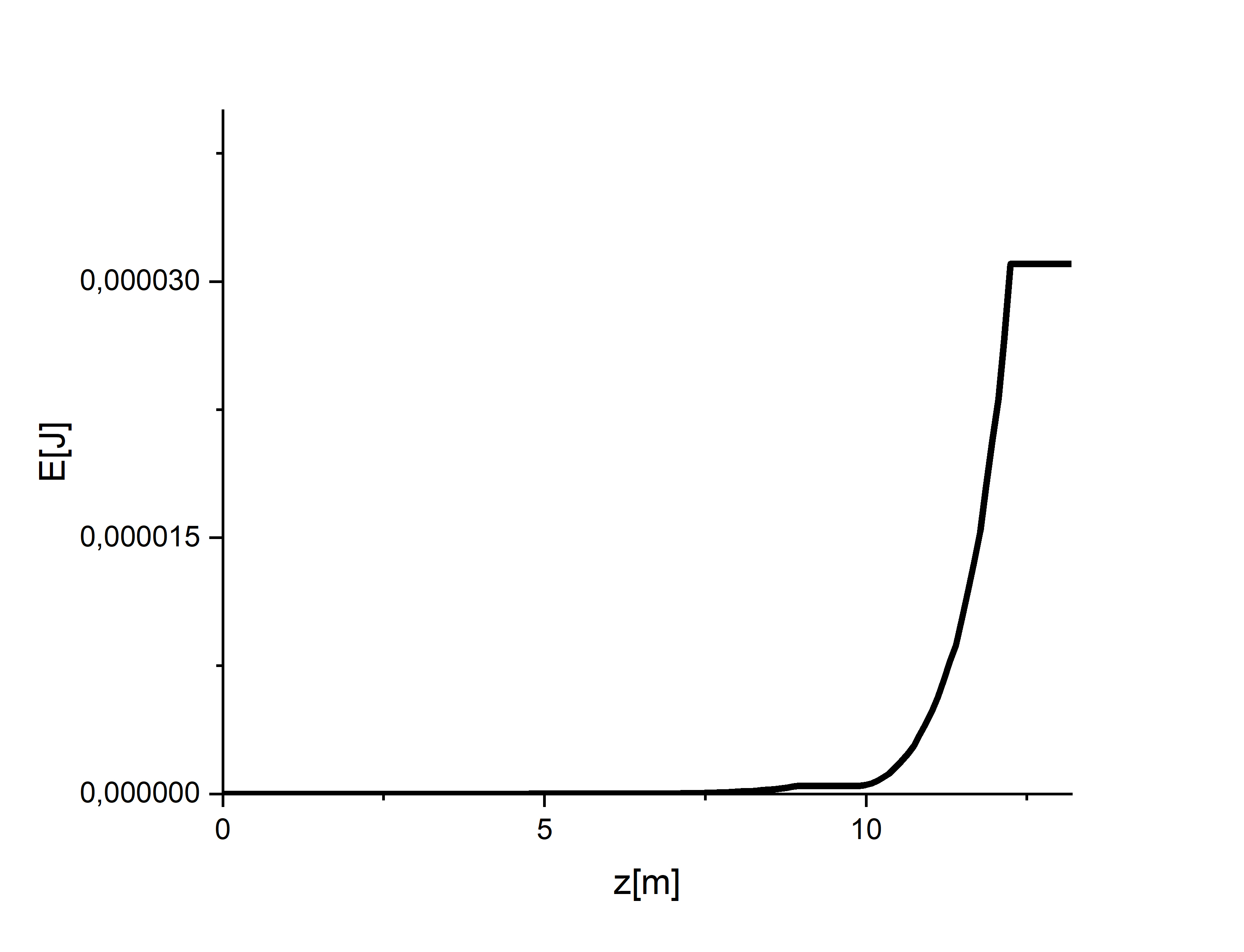}
	\caption{Linear mode of the EUV FEL operation. Evolution of the energy per pulse along the four undulator modules.} \label{Graph5}
\end{figure}

\begin{figure}[tb]
	\centering
	\includegraphics[width=0.7\textwidth]{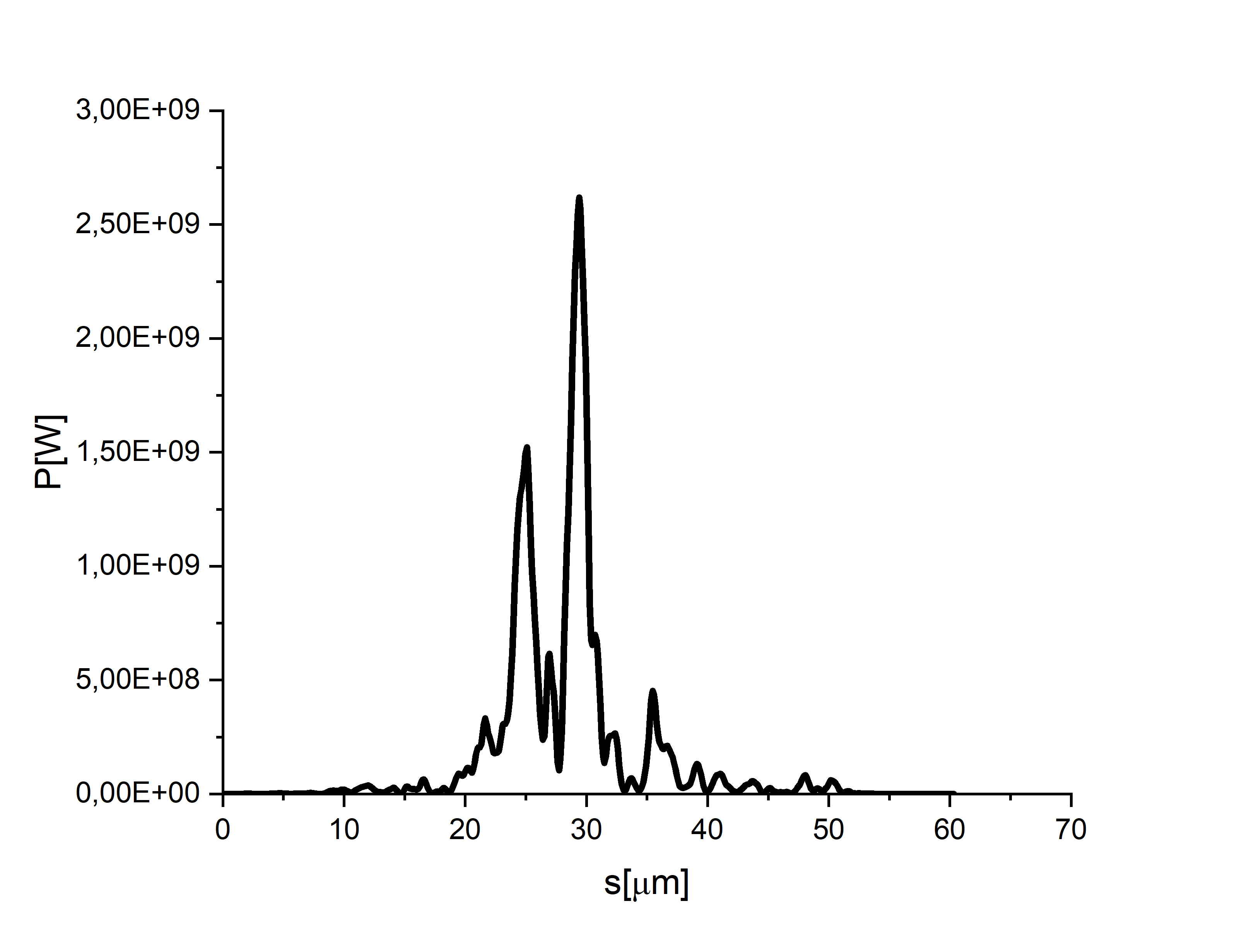}
	\caption{SASE radiation power at the exit of 4th undulator module.} \label{Pw1}
\end{figure}

\begin{figure}[tb]
	\centering
	\includegraphics[width=0.7\textwidth]{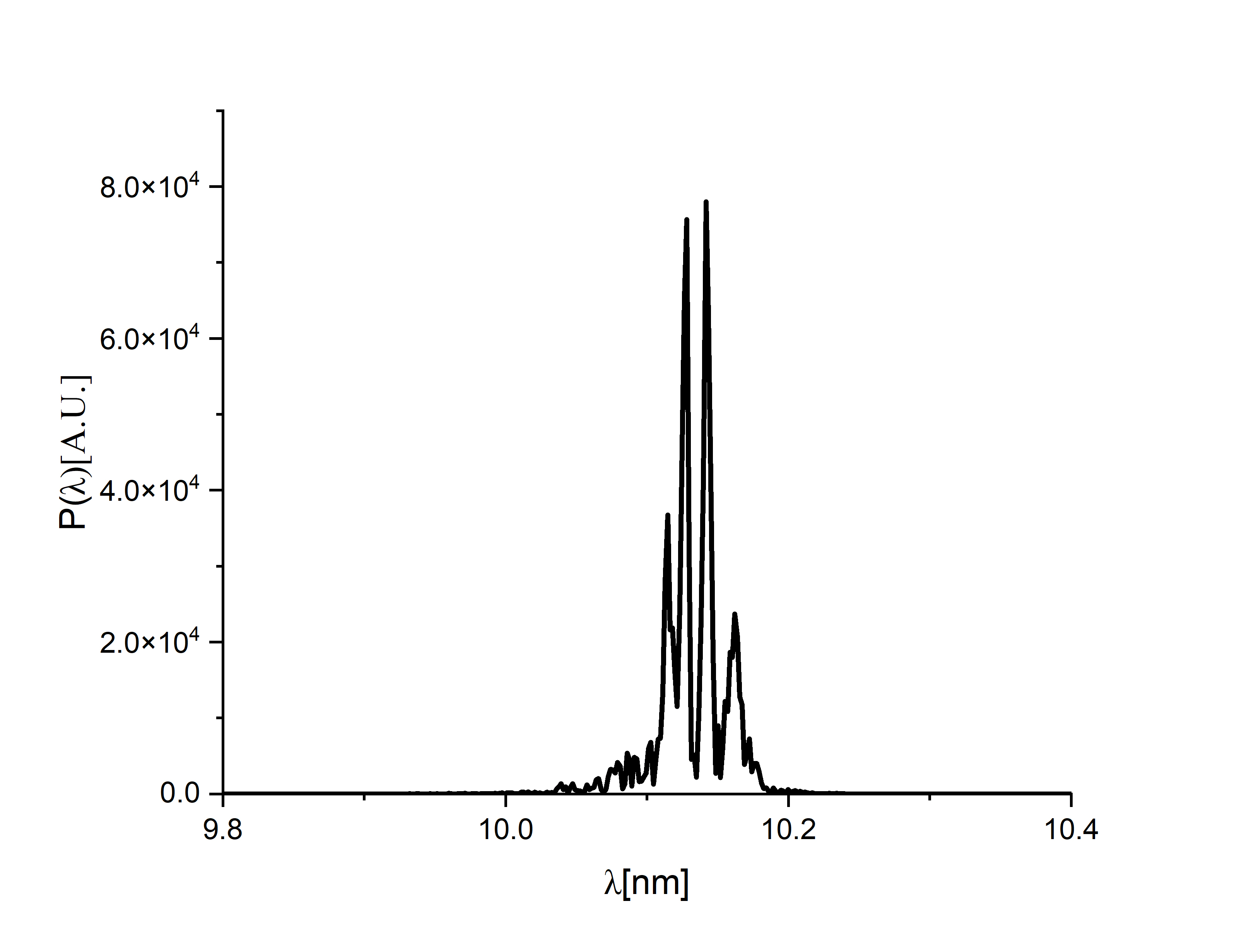}
	\caption{Linear mode of the EUV FEL operation. SASE spectrum at the exit of 4th undulator module.} \label{Sp1}
\end{figure}

\begin{figure}[tb]
	\centering
	\includegraphics[width=0.7\textwidth]{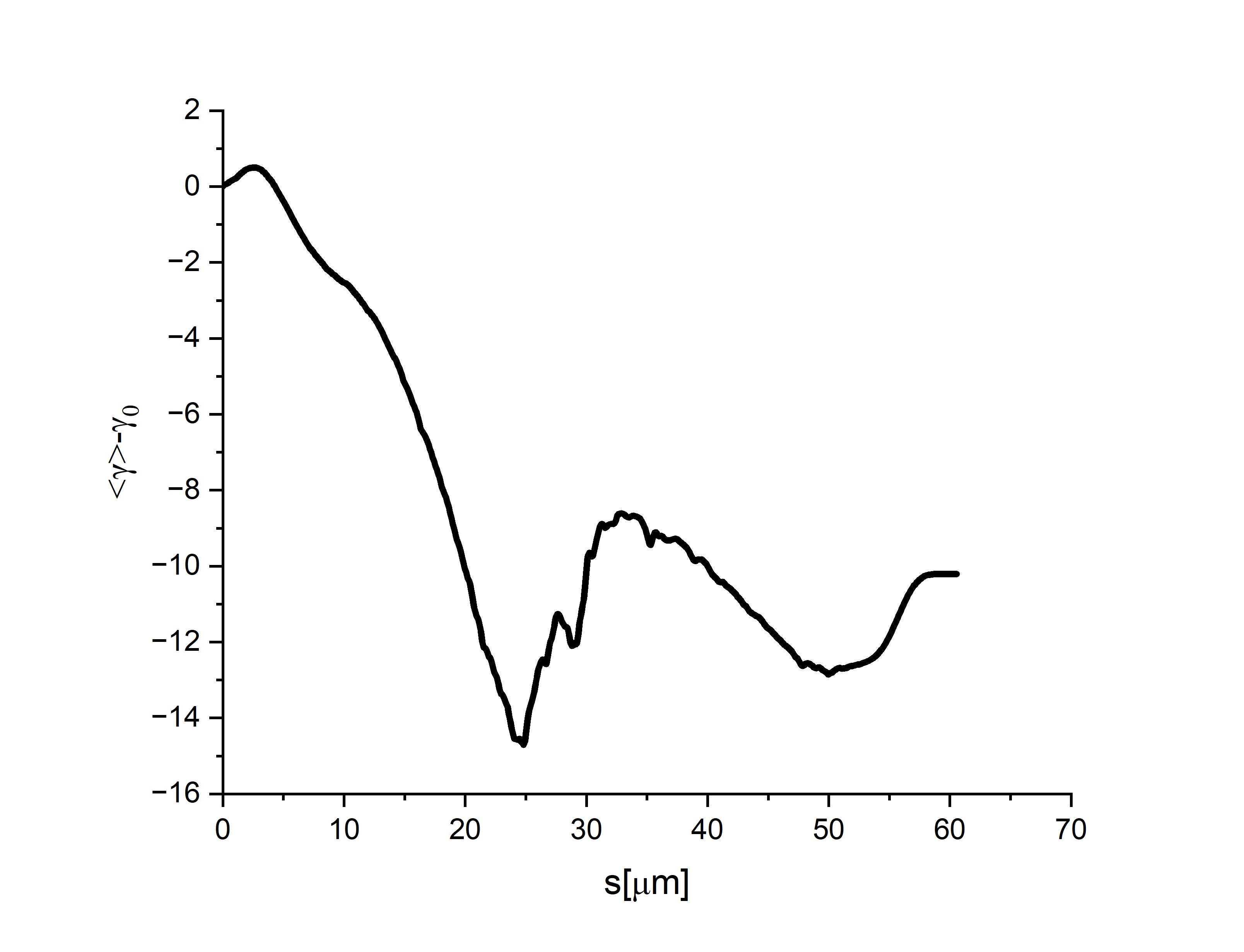}
	\caption{Linear mode of the EUV FEL operation. Energy profile along the bunch at the exit of 4th undulator module.} \label{Eng1}
\end{figure}

\begin{figure}[tb]
	\centering
	\includegraphics[width=0.7\textwidth]{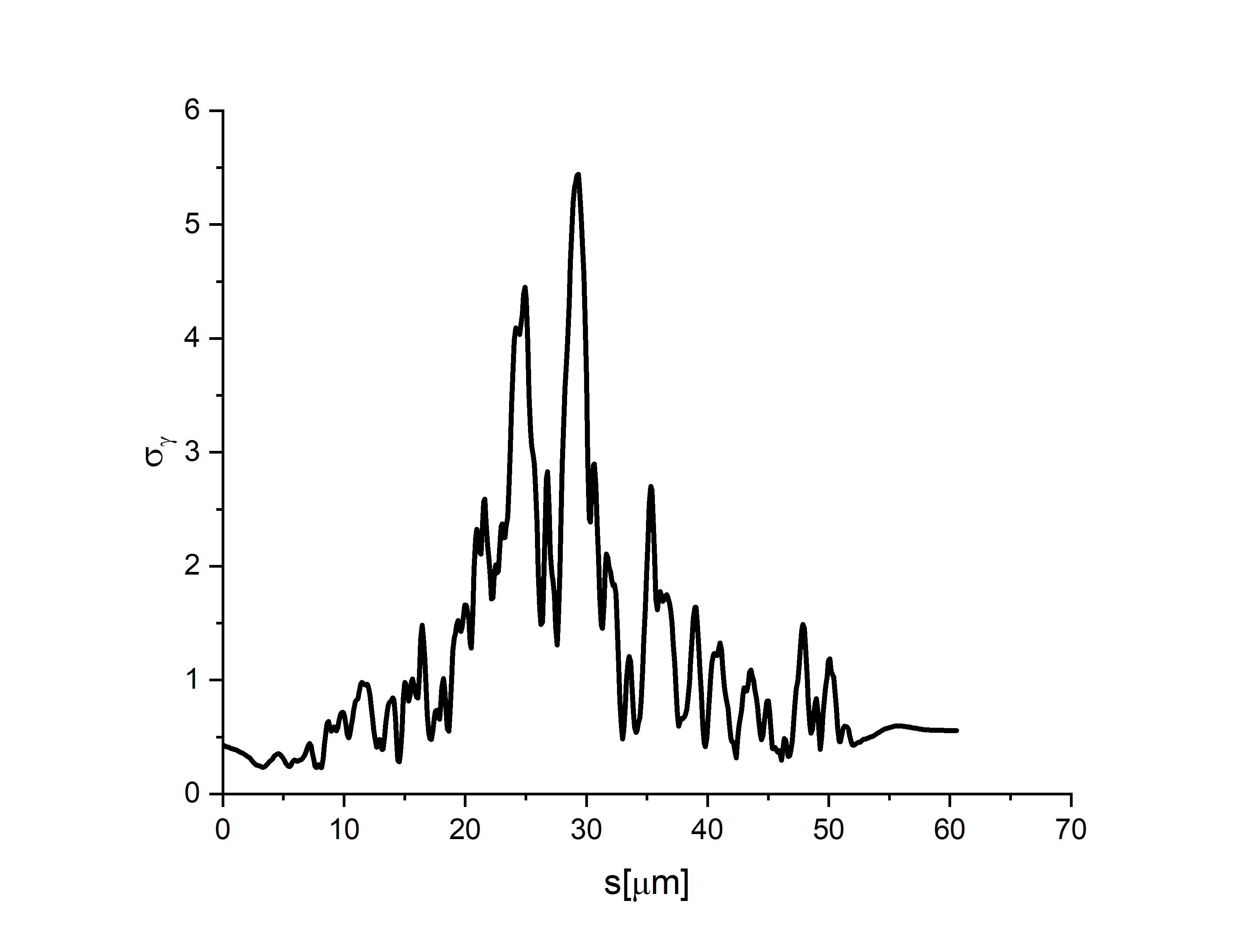}
	\caption{Linear mode of the EUV FEL operation. Electron beam energy spread profile at the exit of the 4th undulator module.} \label{EngSp1}
\end{figure}

\begin{figure}[tb]
	\centering
	\includegraphics[width=0.7\textwidth]{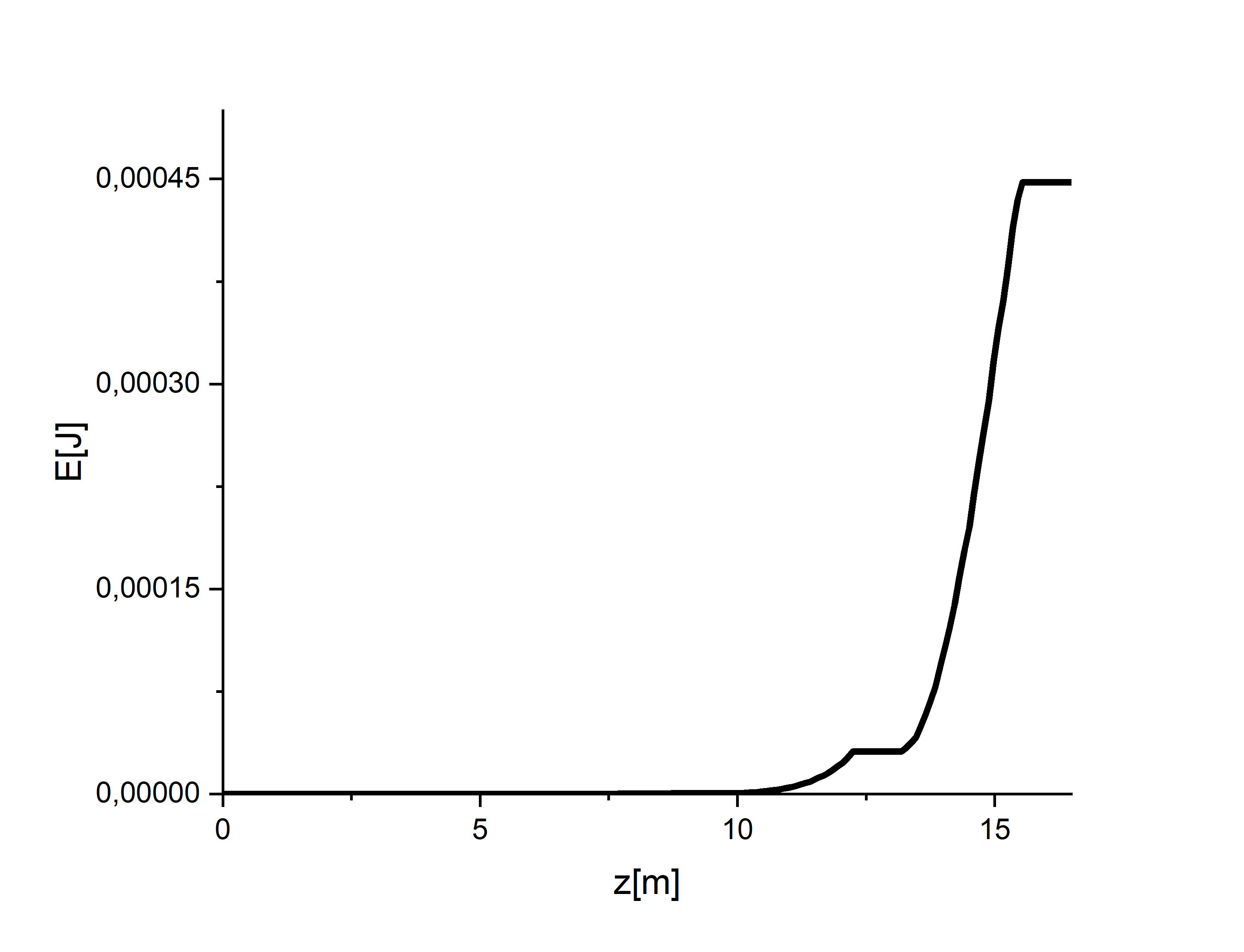}
	\caption{Linear mode of the EUV FEL operation. Evolution of the energy per pulse along the five undulator modules.} \label{Graph4}
\end{figure}

\begin{figure}[tb]
	\centering
	\includegraphics[width=0.7\textwidth]{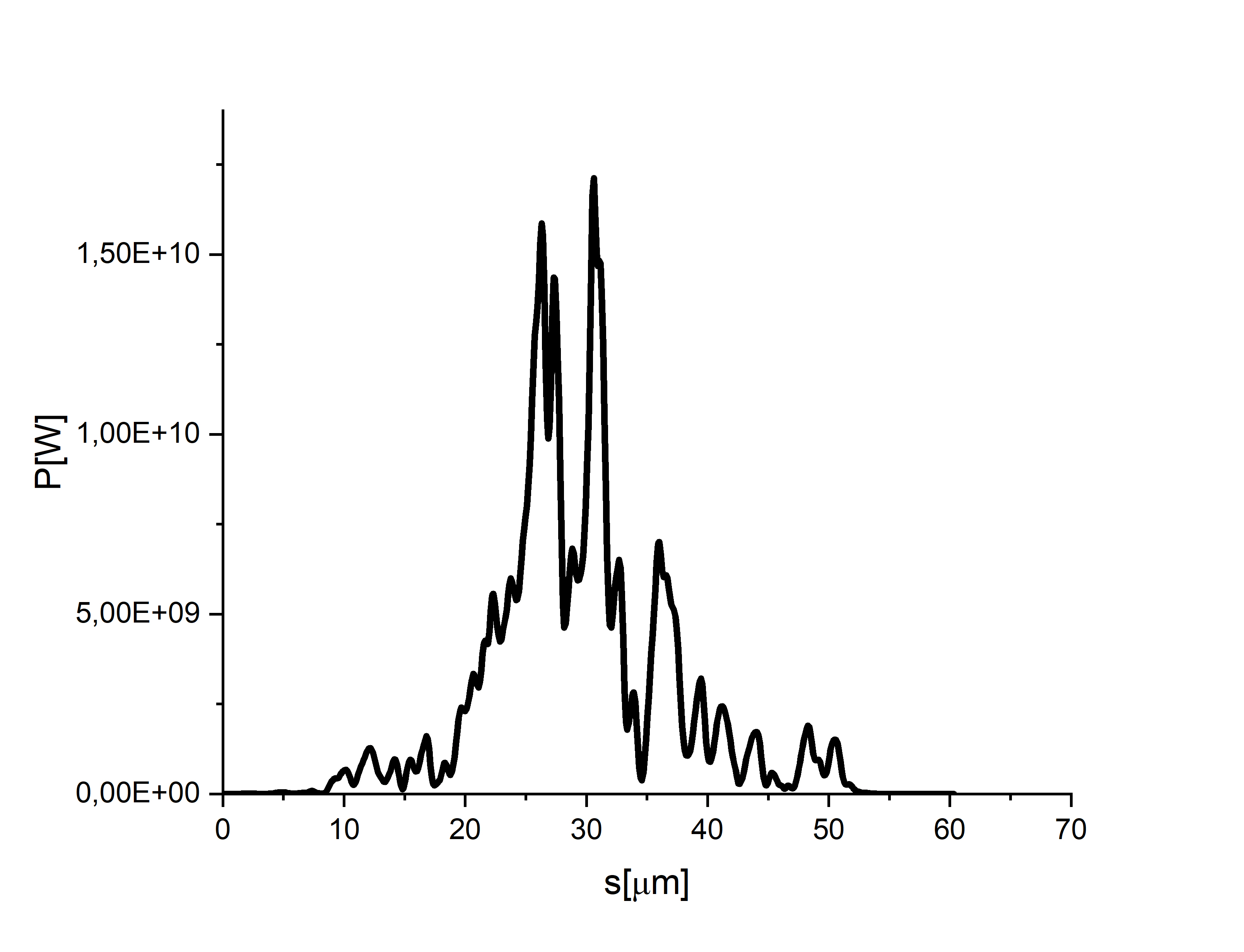}
	\caption{SASE radiation power at the exit of 5th undulator module.} \label{Pw2}
\end{figure}

\begin{figure}[tb]
	\centering
	\includegraphics[width=0.7\textwidth]{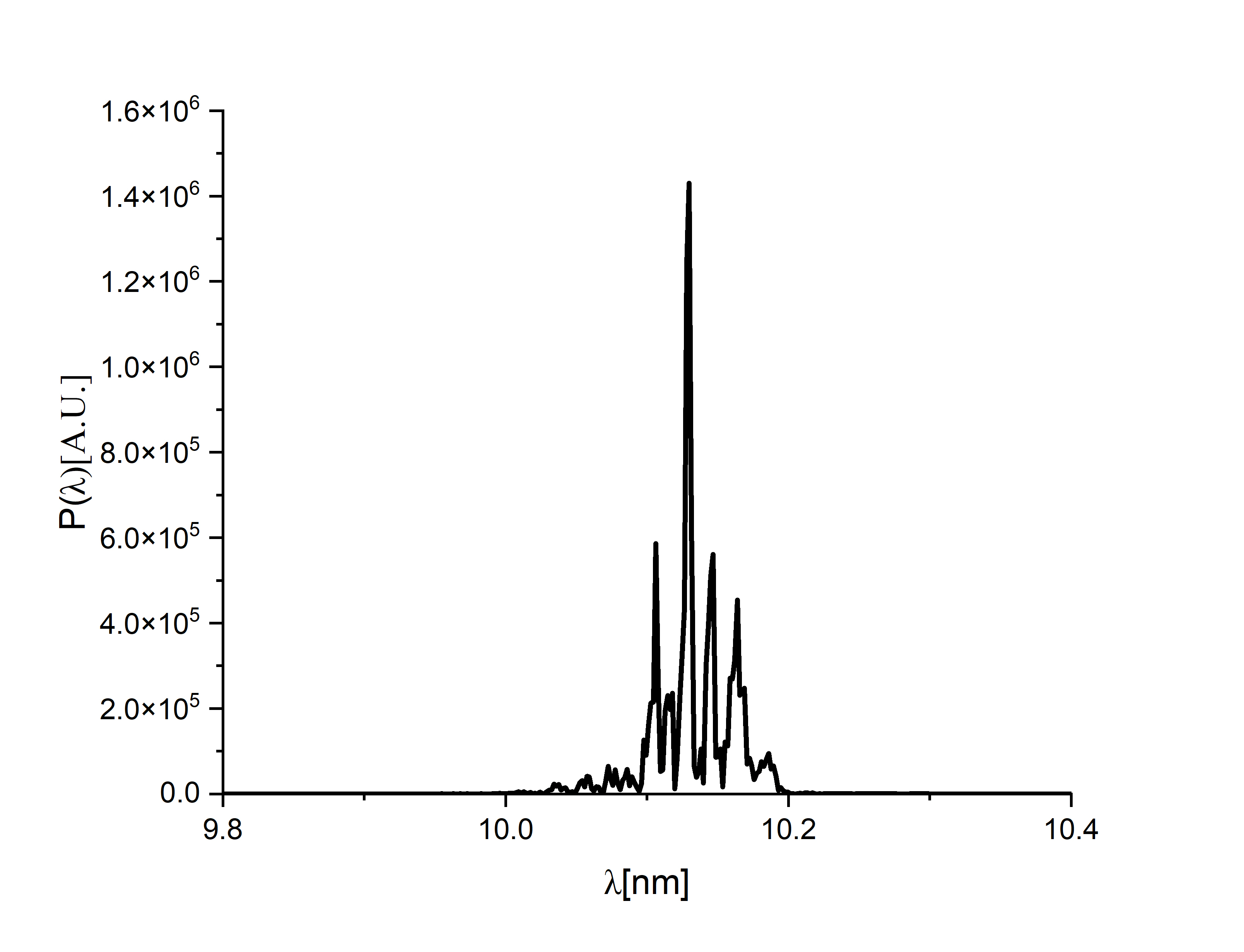}
	\caption{Linear mode of the EUV FEL operation. SASE spectrum at the exit of the 5th undulator module. } \label{Sp2}
\end{figure}

\begin{figure}[tb]
	\centering
	\includegraphics[width=0.7\textwidth]{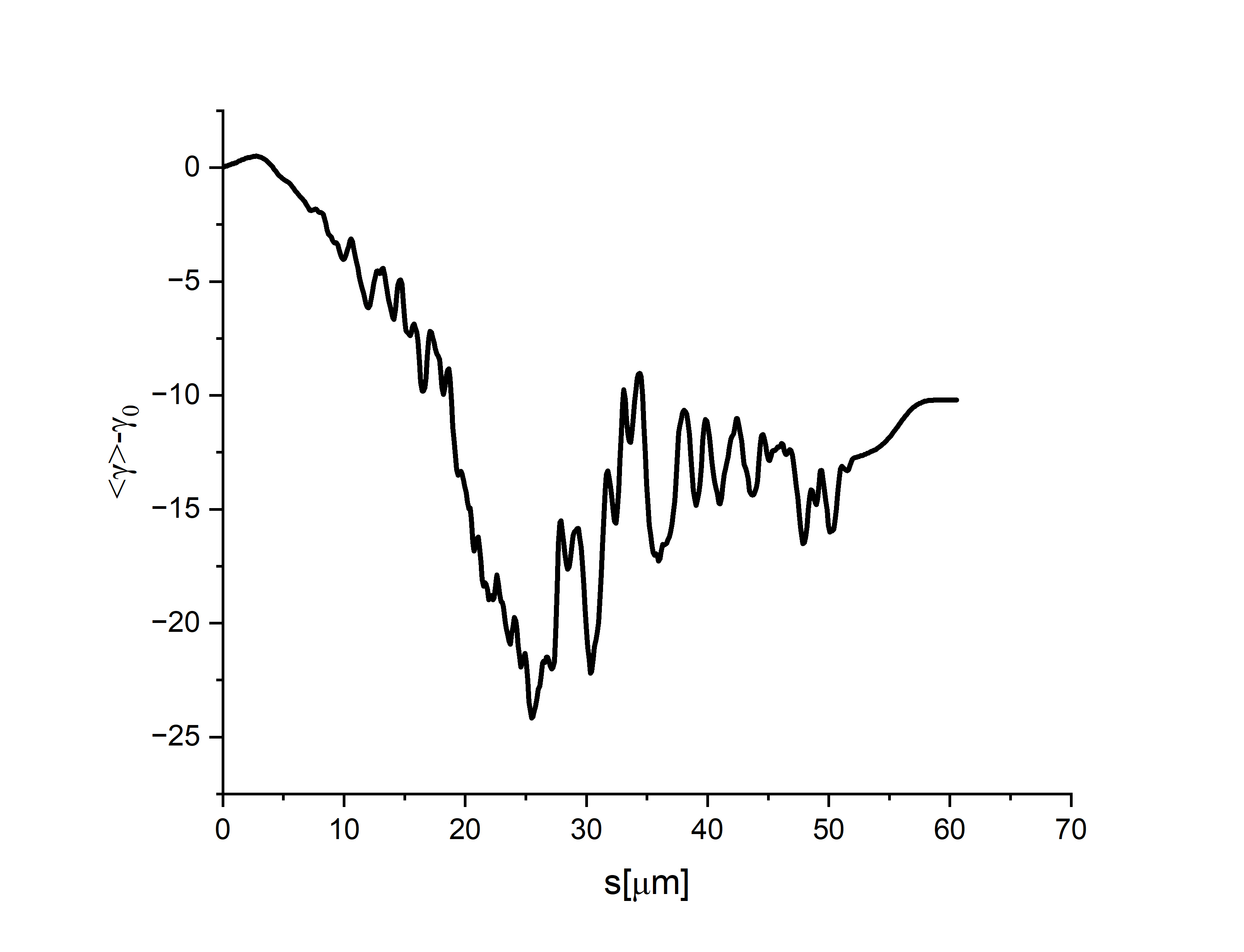}
	\caption{Linear mode of the EUV FEL operation. Energy profile along the bunch at the exit of 5th undulator module.} \label{Eng2}
\end{figure}

\begin{figure}[tb]
	\centering
	\includegraphics[width=0.7\textwidth]{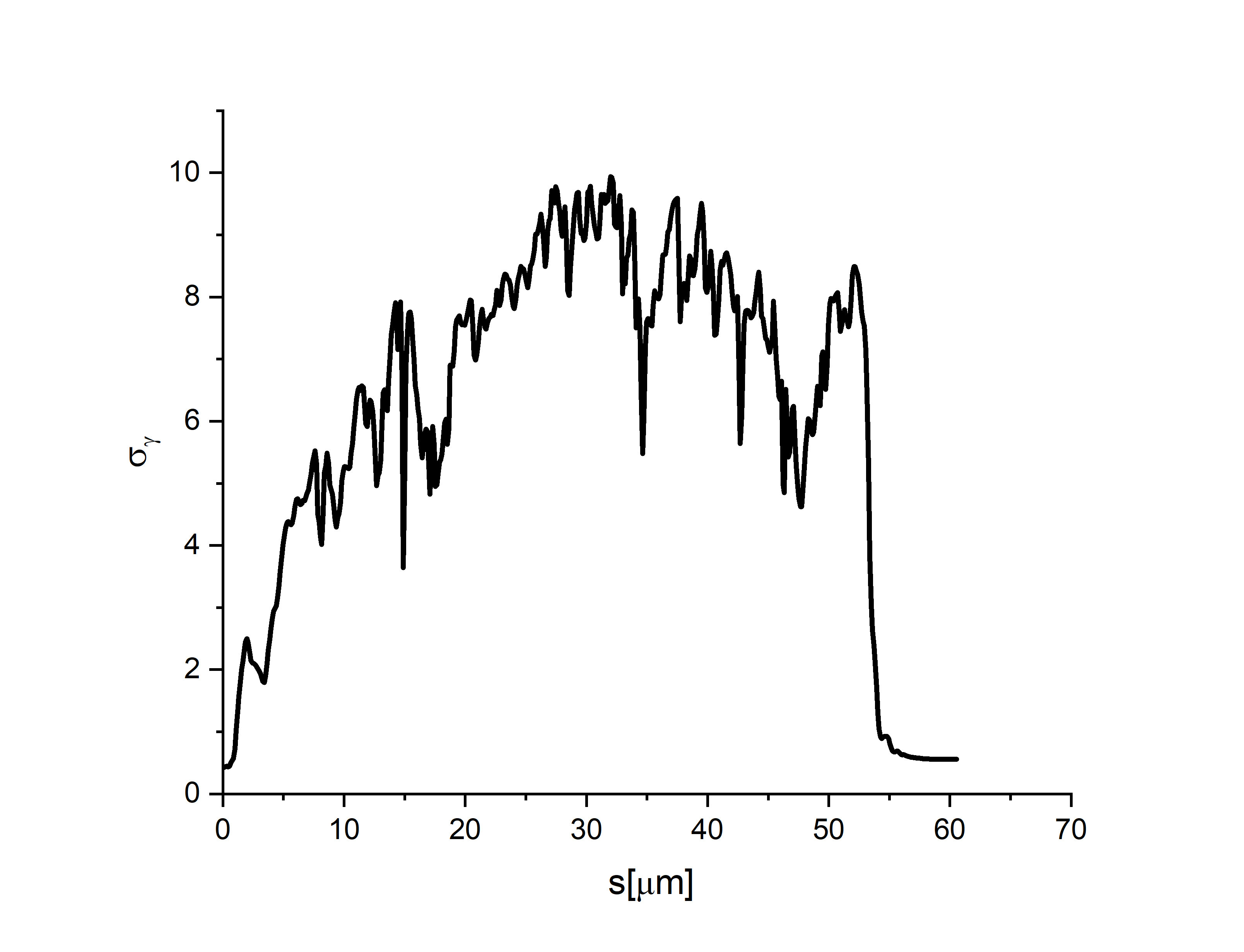}
	\caption{Linear mode of the EUV FEL operation. Electron beam energy spread profile at the exit of the 5th undulator module. } \label{EngSp2}
\end{figure}

The expected beam parameters at the entrance of the FEL undulator obtained in start-to-end simulations \cite{Z} are shown in Fig. \ref{Current} - \ref{EngSp} . As the result of the simulation, we notice a significant initial energy evolution along the electron bunch. In this case, the energy chirp would dominate the electron energy spread, resulting in a wider bandwidth of the X-ray photons within the central cone compared with the calculation in the previous section that accounted for only electron beam energy spread. Therefore, we want to investigate the optimal length (number of sections) of the main undulator.  We performed numerical simulations of the FLASH2 FEL using the code GENESIS 1.3 \cite{G}. The output SASE and electron beam characteristics in the linear regime are shown in Fig. \ref{Graph5} - \ref{EngSp2}. These figures show the spent electron beam parameters obtained for the specific main undulator setting of 10 nm wavelength. The FEL-induced energy chirp comparable to the initial one is achieved at the exit of the 5th undulator module. For the spontaneous photon flux at the exit of the transport photon beamline within the central cone at the 3rd and 5th harmonic, we find about $10^8$  photons/s and $10^7$ photons/s in a $10^{-2}$ bandwidth respectively. The simulations show that this pump-probe operation is possible with FEL pulse energies of about a few hundred mkJ (and consequently a few GW peak power).

\section{Conclusions}

The FLASH2 accelerator complex produces ultra-short electron bunches with a duration approaching sub-hundred femtoseconds. It is natural to take advantage of these ultra-short bunches to provide X-ray radiation. The X-ray radiation pulses can be generated by the spent electron beam downstream of the EUV undulator. The design of our X-ray source is based on the exploitation of the APPLE-III undulator (recently implemented at FLASH2) in the spontaneous emission mode of operation. This concept of using the spent electron beam provides femtosecond X-ray pulses that are naturally synchronized to the EUV SASE FEL laser. 
The proposed femtosecond X-ray source for EUV pump/X-ray probe experiments at FLASH2 is compatible with the FLASH2 layout and can be realized without modifications to the baseline setup. The transport of the X-ray pulse (together with the EUV pulse) to the experimental hall can be performed by exploiting  the existing EUV radiation beamline without modifications. Thus the presented scheme offers great opportunities for ultrafast EUV pump/X-ray probe experiments as the intrinsic jitter between the X-ray and the EUV pulse is completely eliminated.
 \footnote{Above we considered the EUV FEL pulse pump/X-ray probe mode of operation. 
It should be noted, however, that there is another possibility to exploit the sub-hundred femtosecond electron bunch duration at the FLASH2 facility for time-resolved X-ray absorption spectroscopy. This is a well-known (conventional) optical laser pump/X-ray probe scheme. The optical laser system (for optical laser pump/EUV FEL pulse probe experiments) in the FLASH experimental hall provides laser pulses down to 100 fs pulse duration, if necessary in the same bunch train repetition scheme as the electron bunch formation system. The optical laser is synchronized to the radio-frequency source driving the accelerator. A shot-to-shot diagnostic based on an electro-optical sampling system determines the jitter between the optical laser pulse and the electron bunch with an accuracy below 100 fs. Thus,  the temporal resolution decreases from the actual jitter of the accelerator to the accuracy of the jitter measurement \cite{PP}.}.

\section{Acknowledgments}

 We thank Marcus Tischer and  Rolf Treusch for useful discussions.

\newpage


\end{document}